\documentstyle[11pt,twoside,epsf,epsfig]{article}
\begin{document}

\title{The cannonball model of gamma ray bursts}
 \author{Arnon Dar\\
 Technion, Israel Institute of Technology, Haifa 32000, Israel}

\maketitle
\begin{abstract}

The cannonball model (CB) of gamma ray bursts (GRBs) is incredibly more
successful than the standard blast-wave models (SM) of GRBs, which suffer
from profound inadequacies and limited predictive power. The CB model is
falsifiable in its hypothesis and results. Its predictions are summarized
in simple analytical expressions, derived, in fair approximations, from
first principles. It provides a good description on a universal basis of
the properties of long-duration GRBs and of their afterglows (AGs).

\end{abstract}

\noindent
{\bf The CB model} of GRBs [1,2,3,4] assumes that bipolar jets of
highly relativistic CBs are launched axially in core-collapse supernova
explosions (SNe). The CBs are assumed to be made of ordinary matter, as
suggested by the emission of Doppler-shifted lines from the jetted CBs,
ejected by the mildly relativistic $\mu$-quasar SS 433. Crossing the SN
shell (SNS) and the wind ejecta from the SN progenitor with a large
Lorentz factor, the front surface of a CB is collisionally heated to keV
temperatures. The quasi-thermal radiation it emits when it becomes
visible, boosted and collimated by its highly relativistic motion, is a
single $\gamma$-ray pulse in a GRB. The cadence of pulses reflects the
chaotic accretion and is not predictable, but the individual-pulse
temporal and spectral properties are predictable [2].
The GRB afterglow is mainly synchrotron radiation from the
electrons that the CBs gather when they continue their voyage through the
interstellar medium (ISM). It is blended with the light from the host
galaxy and their smoking gun -- the SNe.

\noindent
{\bf Jetted CBs vs collimated fireballs.} Radio, optical and X-ray
observations with high spatial resolution show that $\mu$-quasars eject
relativistic plasmoids along the axis of their accretion disk when matter
accretes abruptly onto their central compact object. In GRS 1915+105,
the observations are compatible with {\it initial} lateral expansion
with a transverse velocity (in their rest system) comparable to
$c/\sqrt{3}$.  In XTE J1550-564 and in many quasars,
such as Pictor A, the ejecta appear to travel long distances without
significant lateral expansion.

\noindent
In the CB model, the jetted CBs, like those observed in
$\mu$-quasars, are assumed to contain a tangled magnetic field. As they
plough through the ISM, they gather and scatter its
constituent protons.
The re-emitted protons exert an inwards pressure on the CBs which counters
their
expansion. In the approximation of isotropic re-emission in the CB's rest
frame and constant ISM density, $n_p$, one finds
that within a few minutes
of an observer's time, a CB reaches its
asymptotic radius $R$. In the same approximation one may compute the

magnetic field that sustains the inwards pressure of the outgoing protons
and derive an explicit law for the observed deceleration of CBs in the
ISM, which depends on the initial $\gamma=\gamma_0$ as they exit the SNS,
on a ``deceleration'' parameter $x_\infty$
and on their viewing angle, $\theta\, , $ relative to their direction
of motion:
\begin{equation}
{1\over\gamma^3}-{1\over\gamma_0^3}
+3\,\theta^2\,\left[{1\over\gamma}-{1\over\gamma_0}\right]=
{6\,c\, t\over (1+z)\, x_\infty}\, .
\label{cubic}
\end{equation}
CBs decelerate to $\gamma (t)=\gamma_0 / 2$ in a distance
$x_\infty/\gamma_0\, ,$ typically of length ${\cal{O}}( kpc)\, .$
Eq.~(1) describes well the deceleration of CBs observed in XTE J1550-564.

\noindent
The original blast-wave models
assumed that GRBs and their afterglows are produced by spherical
fireballs.
The 1997 discovery of BeppoSAX that GRBs have afterglows that appear to
decline with time like a power-law was generally accepted as
undisputable evidence in support of the model. However, spherical emission
implies implausible energy release from small volumes.
Repeated claims made by us
since 1994 [5] that cosmological GRBs
and their afterglows [6,7] are beamed emissions from highly
relativistic jetted ejecta from stellar collapse, were olympically
ignored.  GRB990123 with its record ``equivalent'' spherical energy
release in observable $\gamma$ rays was the turning point of the spherical
blast
wave models. Fireballs became firecones [8,9]
or, more properly, firetrumpets, jets of material funneled
in a cone with an initial opening angle (also called $\theta$) that
increases as the ejecta encounter the ISM (see
Fig. 1a).
\begin{figure}[htb]
\psfig{file=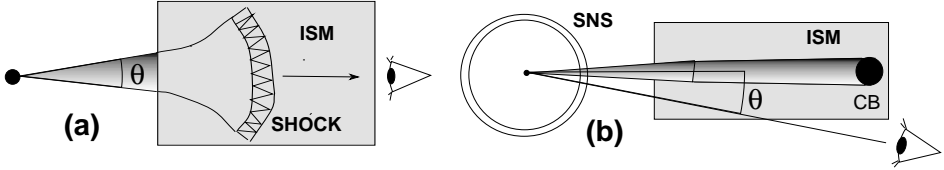,width=12.5cm}
\vspace{-0.3cm} \caption{(a)
Standard-Model geometry. (b) CB-model geometry.}
\end{figure}
For years the modellers, unaware of the Copernican revolution, placed us,
the observers, at a privileged position, precisely on-axis, so that all
detected GRBs would point exactly to us. More recently, the SM view is
evolving (as usual, without proper references) towards the realization
that the observing angle {\it also} matters [10,11,12,13],
a step in the right direction, advocated by the CB model [1-4]: the
observation angle is the {\it only} one that matters.

\noindent
{\bf The GRB -- SN association:}
SNe II/Ib/Ic are far from being standard candles. But if they
are not spherically symmetric  --as they would be if a fair fraction of
them emitted
bipolar jets of cannonballs-- much of their diversity can be due to the
angle from which we see them.
Exploiting this possibility to its extreme, i.e., using SN1998bw as an
ansatz standard candle, Dar suggested [14,15] that all GRB
afterglows should contain a contribution from an SN1998bw placed at the
GRB position.  Dar and De R\'ujula [1] and Dado et al. [3]  have
shown that the optical AG of {\it all} GRBs with known
redshift $ z<1.12$) contain either evidence for an
SN1998bw-like contribution to their optical AG (GRBs 980425, 970228,
990712 and 991208; the first and last one are shown in Fig. 2) or
clear hints in the cases of GRBs 970508, 980703 and 000418 where
scarcity of data, lack of spectral information and multi-colour
photometry
and uncertain extinction in the host galaxy prevented a firm
conclusion.  This suggested that most --and perhaps all-- of the
long-duration GRBs are associated with 1998bw-like SNe
(in the more
distant GRBs, the ansatz standard candle could not be seen, and it was
not seen).
\begin{figure}[htb]
\vspace{-0.1cm}
\centering
\psfig{file=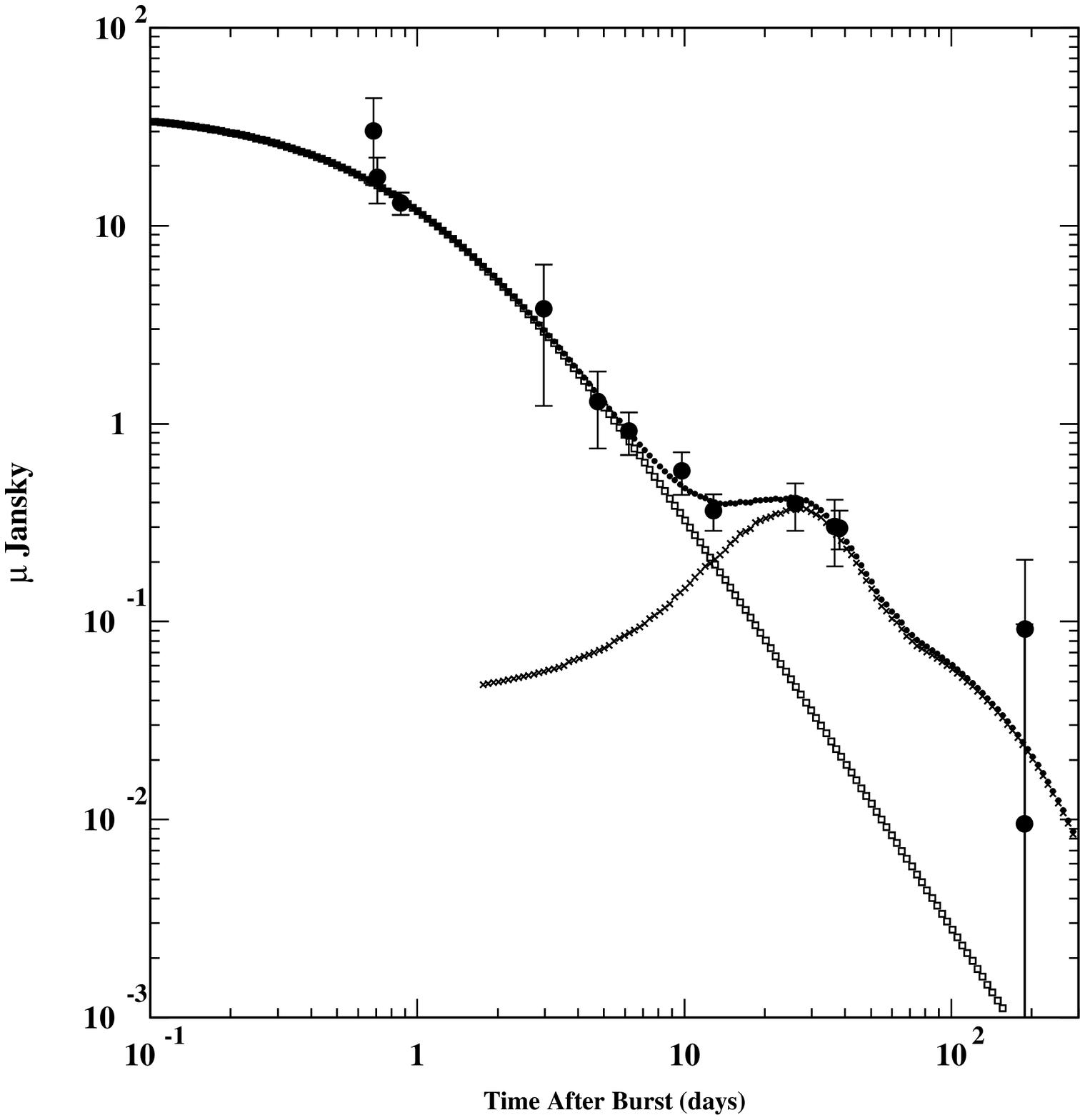,width=6.6cm}\psfig{file=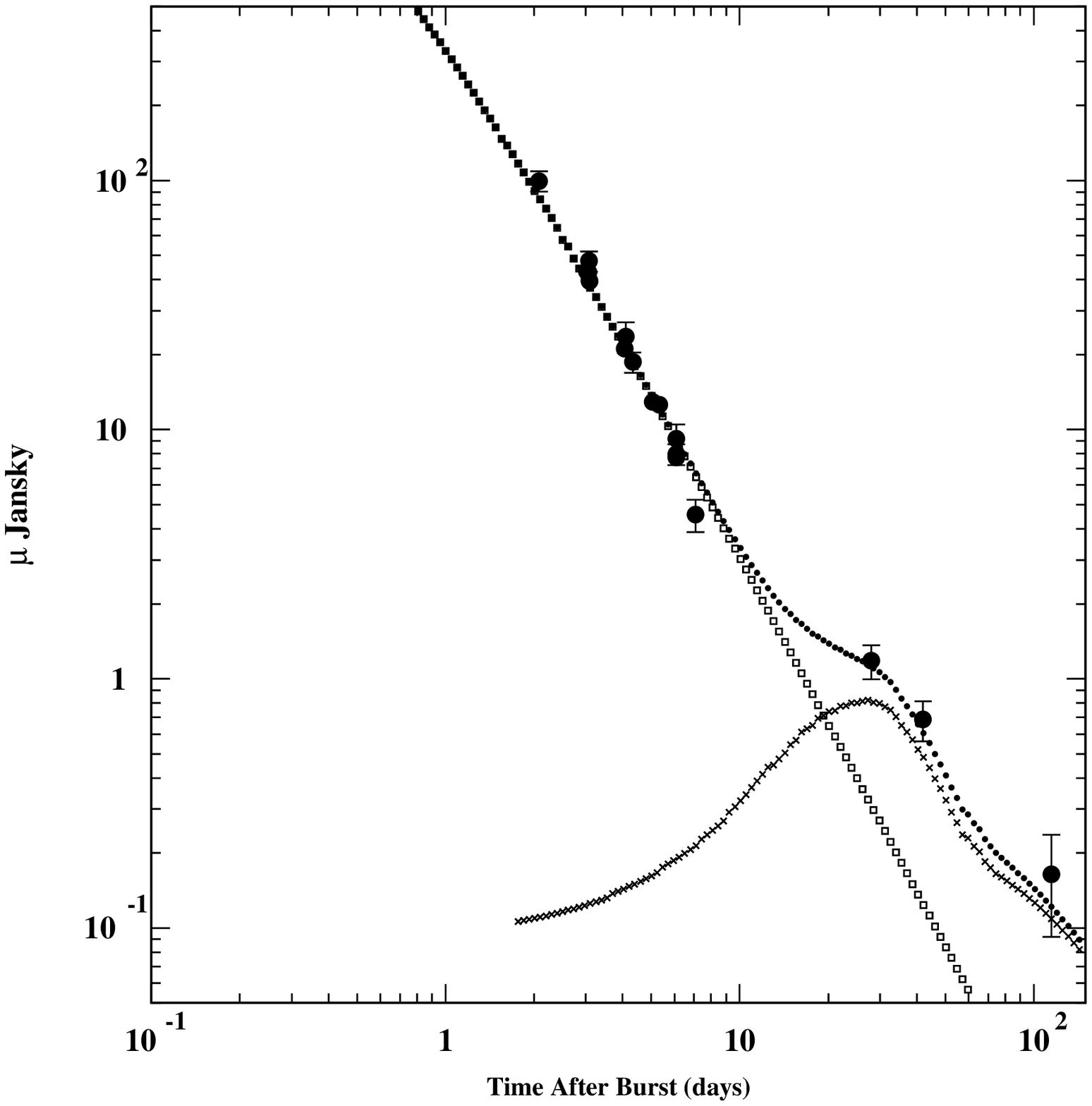,width=6.6cm}
\caption{(a) The R-band AG of GRB970228 with the host-galaxy's
contribution subtracted. (b)
The R-band AG of GRB991208 with the host-galaxy's
contribution subtracted.}
\end{figure}

\noindent
Naturally, ``standard candles'' do not exist, but taking SN1998bw as
a standard candle did
a good job and  gave us enough confidence to {\it predict} how the
associated SN would appear in other GRBs
before they were measured. For instance, in GRB 011121 we used
the first  2 days R-band data to fit the parameters describing the CBs'
contribution
to the AG, to predict explicitly how the AG would evolve with time, and to
conclude [16]
that despite the extinction in the host galaxy {\it ``the SN will
tower in all bands over the CB's declining light curve around day $\sim
30$ after burst''}. The comparison with the data [17],
gathered later, is shown in Fig. 3.
\begin{figure}[htb]
\vspace{-0.1cm}
\centering
\psfig{file=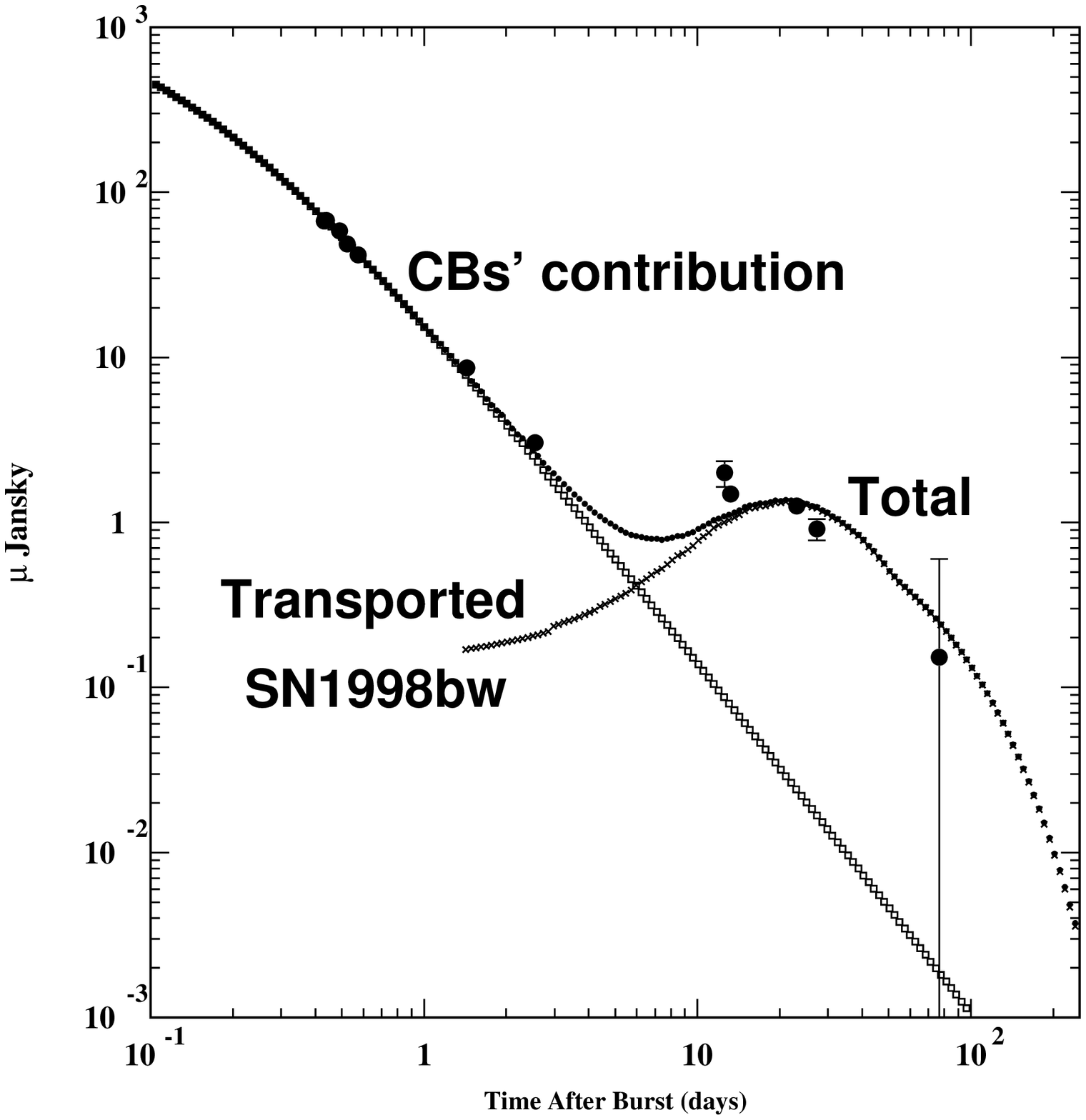,width=6.6cm}\psfig{file=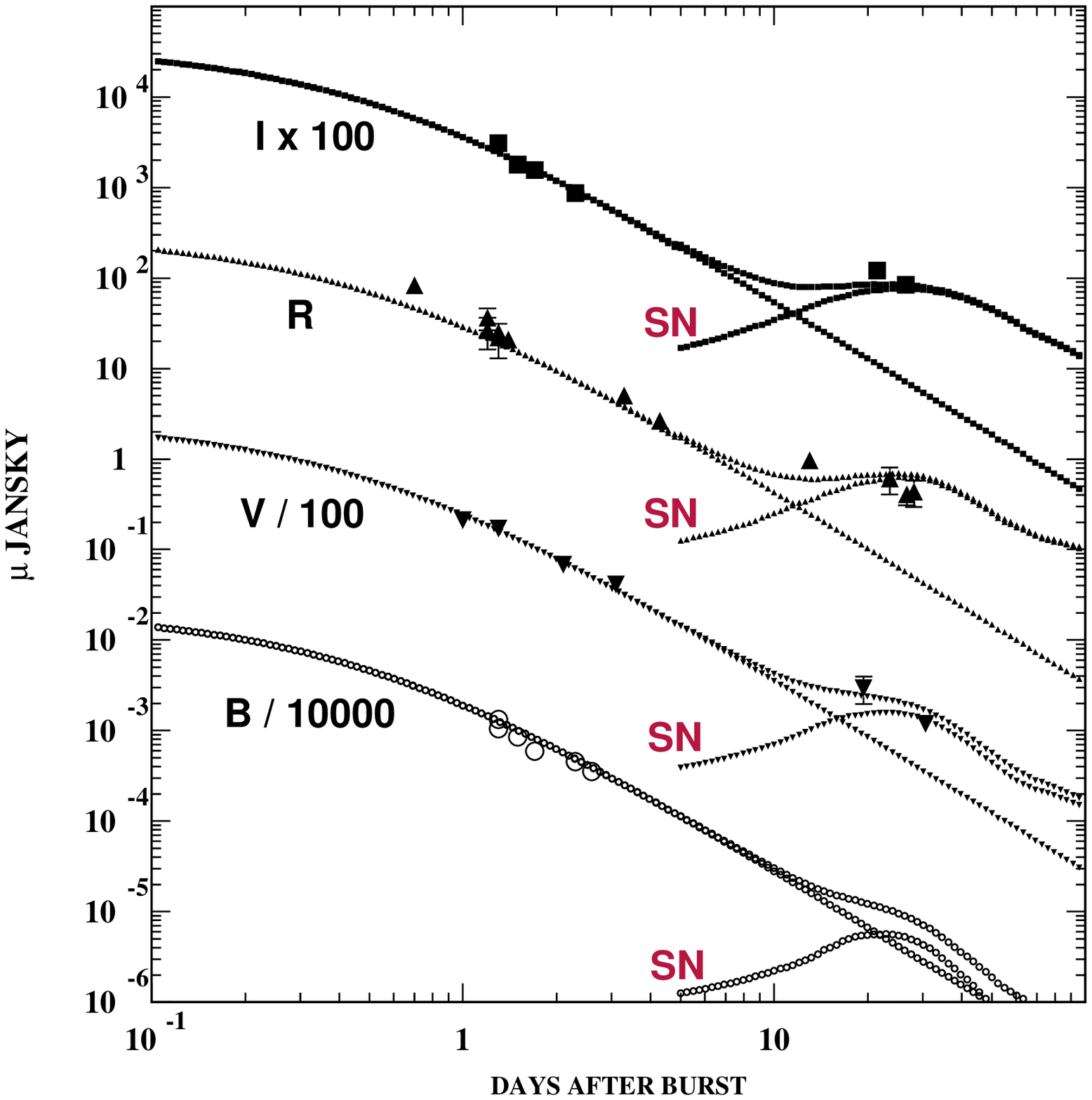,width=6.6cm}
\caption{(a)The R-band AG of GRB011121 with the host-galaxy's
contribution subtracted. (b) Broad-band optical AG of GRB020405
with the host-galaxy's contribution subtracted.}
\end{figure}
The SN spectrum is slightly bluer than that of 1998bw, but not
significantly so. The same exercise was repeated for GRB020405. The very
satisfactory results [18] are shown in Fig. 3b.

\noindent
From our analysis of GRBs and their AGs we deduced that the observed
CBs have typical Lorentz factors $\gamma\sim 10^3$ and are only observable
for angles $\theta$ (between the jet axis and the observer) of
${\cal{O}}(10^{-3})$. For such a small viewing angle, the universal rate
of GRBs and that of core-collapse SNe are comparable, i.e.,  {\it
a good fraction of core collapse SNe emit GRBs}.

\noindent
{\bf The GRB proper.}
During the GRB phase, the CBs are still dense and highly opaque to
protons.  In their rest frame, the rate of energy deposition by the
incident protons near their front face is $ \pi\,R^2\,n_p\, m_p\,
c^3\,\gamma^2\, , $ where R is their radius and $ n_p$ is the
circumburst baryon density.  Approximately, a fraction 1/3 of this energy
that does not escape in neutrinos from $\pi^{\pm}$ decay, is radiated
away. It is Doppler boosted and collimated by the relativistic motion of
the CB, attenuated by the column density between the CB and the distant
observer and redshifted by the cosmic expansion.  For a typical wind
profile, $ n_p \sim r^{-2} $, equilibrium between energy deposition and
radiation implies that an observer sees a surface radiation whose
intensity per unit area is proportional to
\begin{equation}
   I(t) \propto  n_p\, e^{-\sigma_\gamma \int_r^\infty \, n_p\, dr'}
     \sim (t_m/t)^2\, e^{-2\, t_m/t} \, ,
\label{tprof}
\end{equation}
where the observer's time t and the distance r of the CB from the SN are
related through $ dr=\gamma\, \delta\, c\, dt/ (1+z) $ and where the
variation of the Lorentz and Doppler factors of the CB, $\gamma$ and
$\delta$,
respectively, during the short GRB pulse were neglected.
Eq.~(2) has a ``FRED'' shape (fast rise and exponential
decline) with a maximum at $ t=t_m$ when the optical depth to the
observer is $ \tau=2\, . $ The
photons' attenuation cross section is a sum
of the bound-free (bf)
and the Klein --Nishina (KN) cross sections, $ \sigma_\gamma=
\sigma_{bf}+\sigma_{KN}$, at photon energy $
(1+z)\, E_\gamma\, .$ Its energy
dependence produces a ``time lag'' in $ t_m\propto (1+z)/(\gamma\,
\delta\, \sigma_\gamma)\, ,$ which depends moderately on energy
($ \sim E^{-0.3}$) for $ E_\gamma > 4/(1+z)\, keV$ (assuming a solar
composition), but increases rapidly ($ t_m\sim E_\gamma^{-3}$) when
$ E_\gamma$ decreases below $\sim  4/(1+z)\, keV.$ Eq. (2) can
be generalized to other wind profiles.

\noindent
If the energy
deposition rate is balanced by a black-body-like emission, then the
effective  surface  temperature
of the CB seen by a distant observer (Doppler-shifted by a factor
$\delta$ and  redshifted by a factor 1+z  is
\begin{equation}
 T(t) \approx {\delta \over 1+z}\, \left [{m_p\, c^2 \over
                3\, \sigma\, \sigma_\gamma}\right]^{1/4}\, \left [
                 {(1+z)\, \gamma\, t_m\over
                 t^2\, \delta}\right]^{1/4}\, .
\label{Tpeak}
\end{equation}
For the typical observed values $z\sim 1$,  $ t_{max}\sim 1\, s$
and $\gamma\approx\delta\sim 10^3\, ,$
deduced from the afterglows of GRBs  with known redshift
[3,4], one obtains that $ T\sim 0.1 \, MeV$ at maximum intensity. This
explains why the typical $\gamma$-ray  energy in a GRB is [20]
$\sim 250\, keV$
(for a black-body radiation, $ \langle E_\gamma\rangle =2.7\,k\,T\sim
270\, keV)\, .$ For a wind profile $ n_p\sim r^{-2}\, ,$
the peak energy declines like $ E_p(t)\propto T \sim t^{-1/2}$
during the pulse, where t is the time elapsed since the beginning of the
pulse (not the GRB) consistent with observations (e.g., [21]).
An example of $\gamma$-ray light curve, assuming a black-body
emission,
is given in Fig. 4a for the single
pulse of GRB980425, the closest GRB of known redshift $z$.

\begin{figure}[htb]
\vspace{-1cm}
\centering
\psfig{file=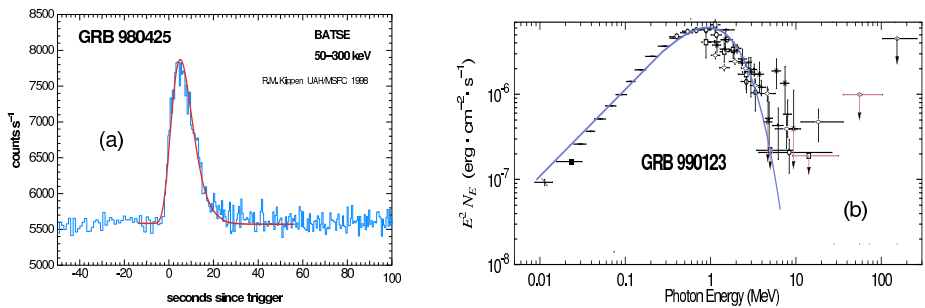,width=12.5cm}
\vspace{-0.3cm}
\caption{(a) Temporal shape of GRB980425. (b) $E^2\,dn_\gamma/dE$
spectrum of GRB990123. }
\vspace{-0.2cm}
\end{figure}
An example of time-integrated
black body emission is given in
Fig. 4b, for the most energetic recorded GRB of known $z$.
The low-energy part of the spectrum, in this
and other GRBs [20], behaves like $ E^2\,dn_\gamma/dE\sim
E^1$, in agreement
with the CB-model's prediction (the SM inescapably
predicts a mean slope disagreeing with observation by $\sim 1/2$ unit
(Ghisellini, these proceedings). The high-energy tail, however, is not
well reproduced by a simplified black-body model [1]; it
should be flatter. This is not surprising: the CB in its rest
system is continuously bombarded
by  particles of high $\gamma$, which
produce via Coulomb interactions a quasi-thermal distribution with
a power-law tail,
$ dn_e/dE\sim E^{-2.2}\, $  instead of an exponential thermal tail.
The bremsstrahlung emission from such a quasi thermal distribution of
electrons from the ablated front face of the CB can be well
interpolated by
\begin{equation}
 E^2\, (dn_\gamma/ dE) \approx  [A\, E^{(p_1-2)s}+
          B\, E^{(p_2-2)s}]^{-1/s}\, ,
\label{spectrum }
\end{equation}
where $ p_1\approx 1$, $ p_2\approx 2.2$, A and B are constants
whose ratio determines the  peak energy and s determines
the sharpness of the transition between the low energy and the high energy
power-law behaviours. Indeed, the observed distributions of $ p_1$  and
$ p_2$ (e.g. [21]) peak at these theoretical values.

\noindent
Because of attenuation, only a fraction $ E_\gamma^T(CB)\sim \pi\,
R^2\,
m_p\, c^2\, \gamma/ 3\, \sigma_\gamma$ of the energy deposited in the CB
is observable. However, in the observer frame,
it is Doppler-boosted and
relativistically collimated to a large $\gamma$-ray
fluence,
\begin{equation}
 F_\gamma [CB]= (1+z)\,\delta^3\, E_\gamma^T(CB)/(\,4\, \pi\, D_L^2)
\label{fluence }
\end{equation}
where $ D_L$ is the luminosity distance of the GRB
and $ \delta\equiv 1/\gamma\,(1-\beta\cos\theta)
\simeq 2\,\gamma(t)/ (1+\theta^2\gamma^2)$
in the domain of interest for GRBs: large $\gamma$ and small $\theta$.

\noindent
A long list of general properties of GRB pulses is reproduced in the
CB-model from these formulae, that unlike the SM have a quasi-thermal
origin (bremsstrahlung as opposed to synchrotron). $
E_\gamma^T(CB)$
inferred from observations, behaves as a standard candle [2,22]
of $ \approx 10^{44}\, erg.$

\noindent
{\bf GRB afterglows.}
Most of the observed SNe take place in super bubbles (SB) of low density.
A CB exiting a SN and the presupernova wind into the low
density SB, soon becomes transparent to its own enclosed
radiation. At that point, it is still expanding and cooling adiabatically
and by bremsstrahlung. If bremsstrahlung dominates the cooling, then
the fluence of the X-ray AG  decreases
with time as $ t^{-5}$ . An example is shown in Fig. 5a.
\begin{figure}[htb]
\vspace{-0.1cm}
\centering
\psfig{file=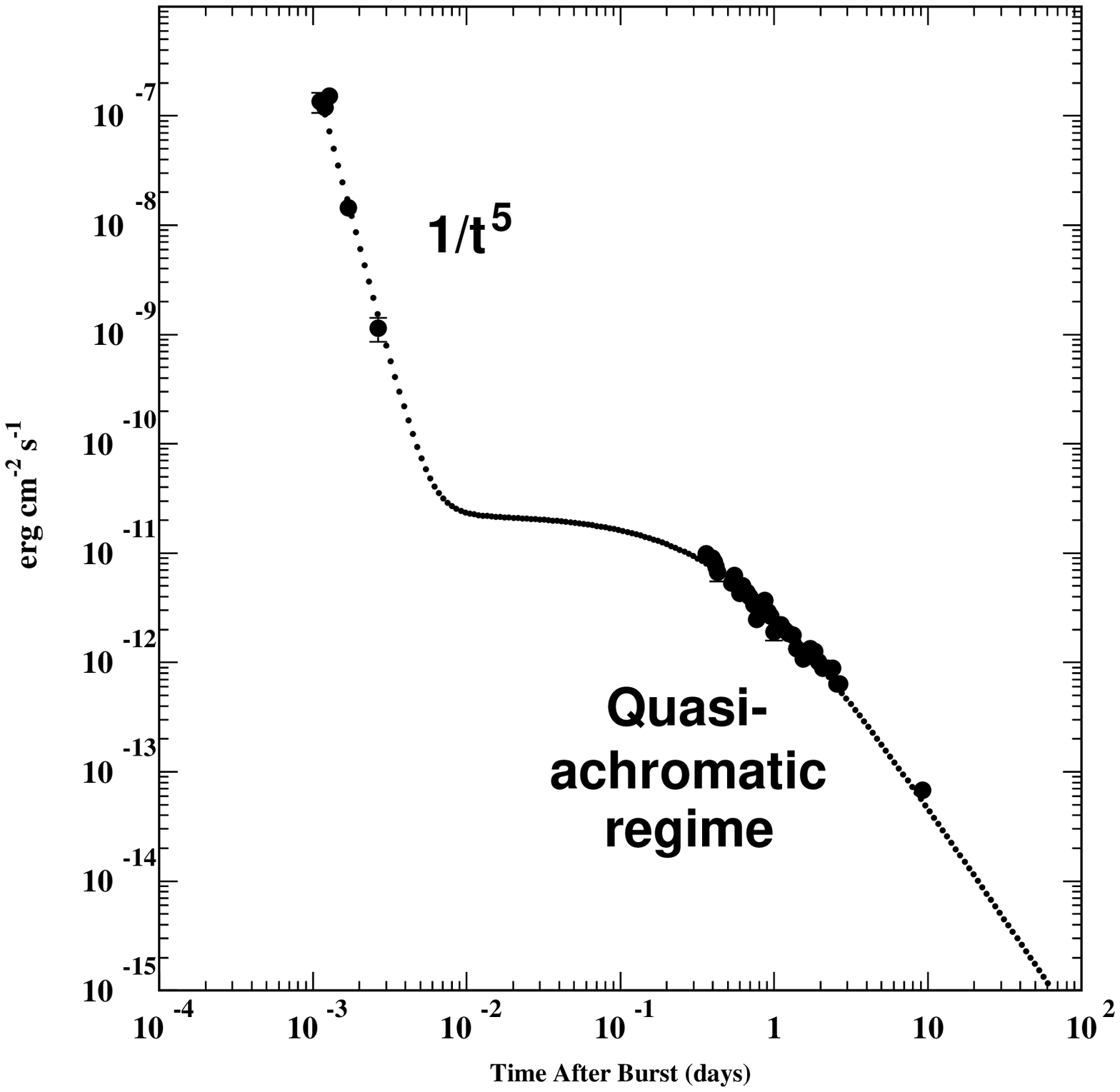,width=6.6cm}\psfig{file=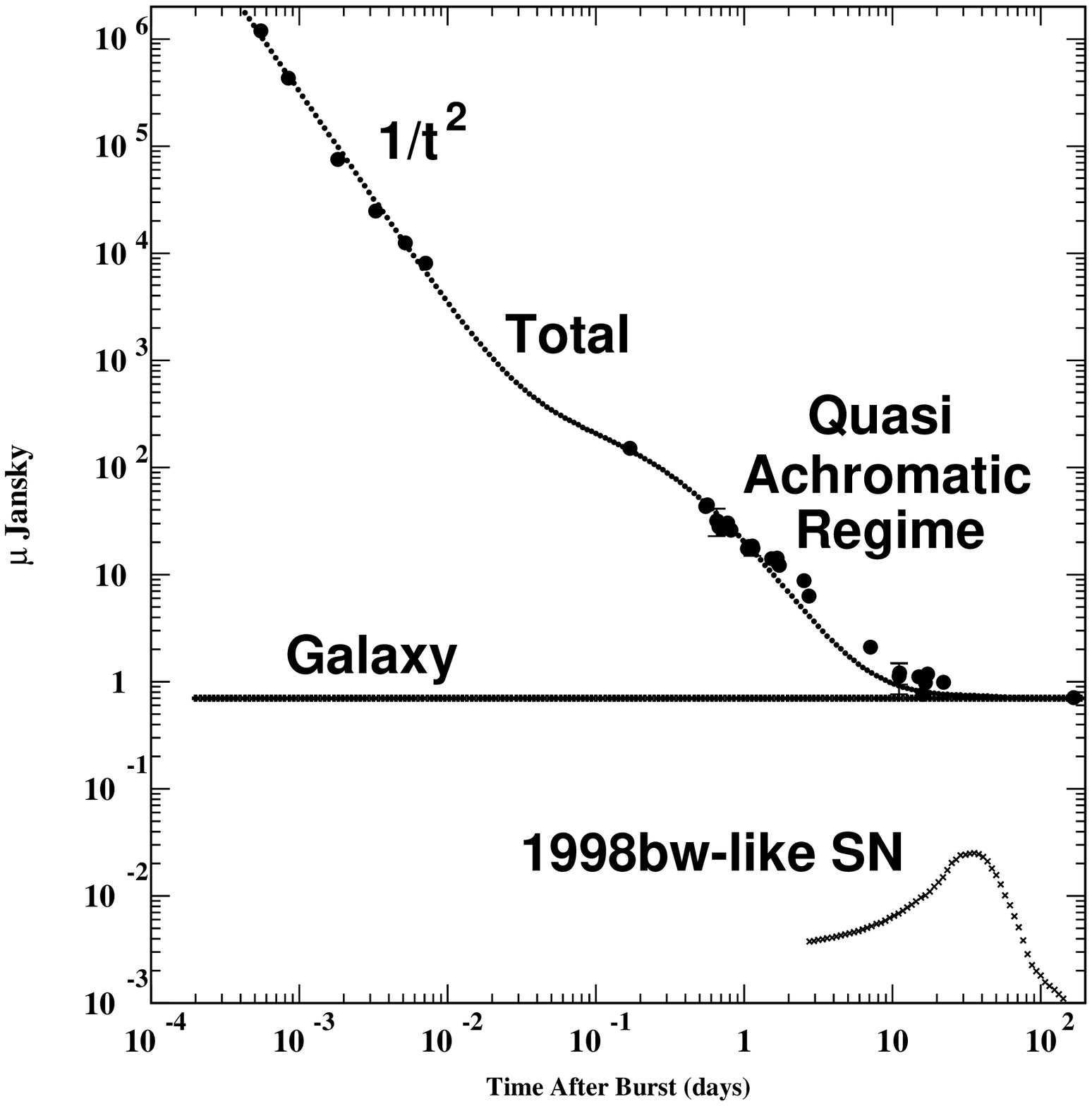,width=6.6cm}
\caption{(a) X-ray AG of GRB010222. (b) R-band AG of GRB990123. }
\end{figure}
Many  X-ray AGs are
compatible with this prediction [3]. If adiabatic
expansion dominates the cooling then the  fluence decreases like
$ t^{-3.5}\, .$ If synchrotron emission takes over while the CB still
propagates  in a $ r^{-2}$ density profile, its fluence
decreases like $t^{-2}$ (see Figs. 5b, 6a).

\noindent
{\bf The optical AGs} of {\it all} GRBs of known $z$  are also well
described in the CB model. They
are synchrotron radiation of the electrons
that the CB gathers in its voyage through the ISM
(line emission and inverse Compton scattering also contribute to the
late X-ray AG). These electrons
are Fermi-accelerated in the CB enclosed magnetic maze and cooled by
synchrotron
radiation to a broken power-law distribution with
an {\it injection break} at the energy
$E_b=m_e\,c^2\,\gamma(t)$
at which they enter the CB.
The emitted synchrotron radiation has a broken
power-law form [4],
with a break frequency corresponding to $ E_b$.
In the observer frame, before absorption corrections, it has the
approximate form:
\begin{equation}
F_\nu \equiv \nu\, (dn_\gamma/ d\,\nu) \propto n_e\, R^2\,
[\gamma(t)]^{3\alpha-1}\, [\delta(t)]^{3+\alpha}\, \nu^{-\alpha}\, ,
\label{sync}
\end{equation}
where  $ \alpha\approx 0.5$ for $\nu\ll\nu_b$
and  $ \alpha\approx p/2 \approx 1.1$ for $\nu\gg\nu_b$, and
\begin{equation}
 \nu_b \simeq 1.87\times 10^3\, [\gamma(t)]^3\,\delta(t)\,
[n_p/ 10^{-3}\;cm^{-3}]^{1/2}/(1+z)\, Hz.
\label{nubend}
\end{equation}
is the ``injection break'' frequency
corresponding to $ E_b $ [3].
Eq. (6) (or the interpolation formula used in [4])
describes well the  observed AGs of all GRBs with known redshift [3,4,]
after subtracting the contribution of the host galaxy and
the SNe from the observed optical AG, and after correcting
for extinction  in our
Galaxy and the host galaxy (which diminishes with time as the CB moves
far away from the explosion site). This is demonstrated in
Figs. 2, 3, 5, and 6.

\noindent
{\bf Temporal breaks.} In the CB model, changes in the temporal
decline rate of the AG of a CB  have two distinct origins:
the deceleration of the CB and changes in the ISM density along its
trajectory:\\
Eq.~(1) implies that $ \gamma(t)$ and consequently also
$ F_\nu(t)\, ,$ change very little when $ t<t_0\, (1+3\,
\gamma_0^2\, \theta^2)\, ,$ where
$ t_0=(1+z)\, x_\infty/6c\, \gamma_0^3\, .$
Later, when $ t \gg t_0\, (1+3\, \gamma_0^2\, \theta^2)$
and $ \gamma^2\, \theta^2\ll 1 $, $ \gamma(t)$ approaches its
asymptotic $\sim t^{-1/3}$ behaviour and
$ F_\nu(t)\sim t^{-(4\alpha+2)/3}\sim t^{-2.13}$ if $\nu\gg \nu_b$
and $\sim t^{-1.33}$ if $\nu\ll \nu_b\, .$
The transition of $F_\nu(t)$
around $ t=t_0\, (1+3\, \theta^2\,\gamma_0^2)$
to its asymptotic behaviour is achromatic and gradual. \\
Eq.~(6) implies that a chromatic break in the AG takes place
when $ \nu_b(t)$ crosses the observed band at time $ t=t_b\, ,$
where $ \nu_b(t_b)=\nu\, .$ If $  F_\nu(t)\sim t^{-\beta}$
before the break, then $  F_\nu(t)\sim t^{-1.6\beta}$ right after it.\\
Eqs.~(1),(6) also imply that variations in the ISM
density induce corresponding variations in $ F_\nu(t)$ and
$ \nu_b(t)\, ,$ which are proportional to $ n_p$ and $ \sqrt{n_p}\, ,$
respectively.

\noindent
All these possibilities have materialized, e.g. in the AG of
GRB021004, [22] as shown in Fig. 7a. Moreover, since the CB model
successfully describes all the observed AGs of GRBs with known redshift,
it does explain also the so-called ``breaks'' in these AGs, when they are
there.  The firetrumpet models have claimed to produce
sharp breaks in the AGs
when the beaming angle becomes larger than
the opening angle of the ejecta [8.9]. However, they are not reproduced by
detailed calculations that properly take into account arrival time and
viewing angle effects. Thus, also the ``standard candle energy'', derived
in the SM [23]
[by extracting opening angles from  temporal ``breaks''
in AGs, is baseless!
\begin{figure}[htb]
\vspace{-0.1cm}
\centering
\psfig{file=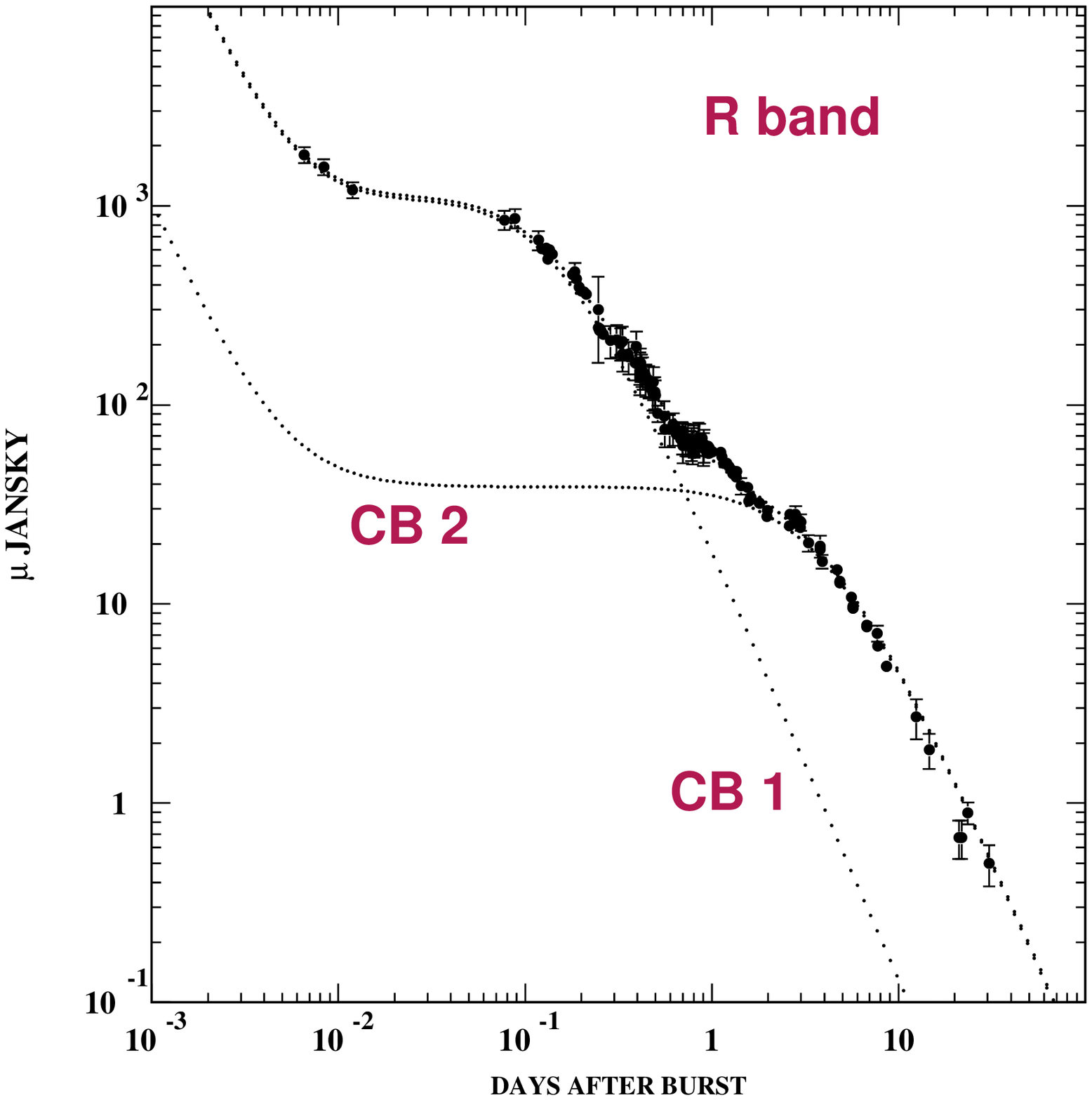,width=6.6cm}\psfig{file=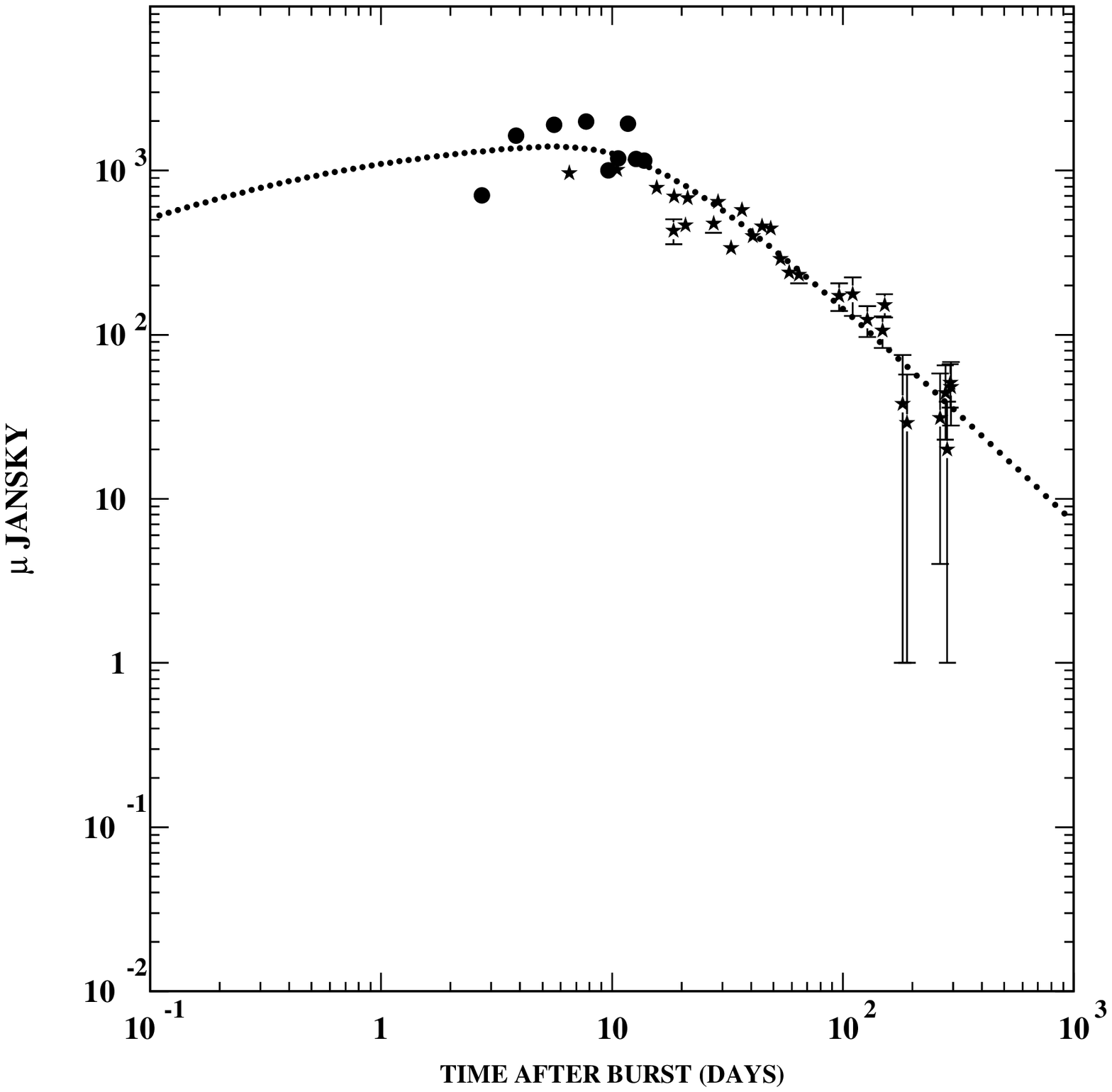,width=6.6cm}
\caption{(a) The predicted host-galaxy-subtracted R-band
light of GRB021004 from a 2 CB broad-band fit to its AG []. (b) The
8.46 GHz radio light curve of GRB991208 and that obtained from the
CB model fit to its broad-band AG}
\end{figure}
Both the temporal and spectral evolution of optical AGs are
well reproduced by the CB model. In particular,
the predicted value, $p\!\approx\! 2.2$, is in agreement with all the data
on X-ray AGs and on relatively late-time optical AGs,  where
$\nu\!\gg\!
\nu_b$ and
$F_\nu\!\propto\! \nu^{-\alpha}$ with $\alpha\!=\!p/2\!\approx\! 1.1$
(after correcting for Galactic extinction).

\noindent
{\bf Broad-band spectra.}
In the radio domain, self-absorption in the CB is important. The dominant
mechanism is free--free attenuation, characterized by a single parameter
$\nu_a$ in the opacity, which behaves as
$\tau_\nu=(\nu_a/\nu)^2\, (\gamma(t)/\gamma_0)^2$.
Absorption in the CB produces  a turn-around of the spectra
from $ F_\nu\sim \nu^{1.5}$ to  $ F_\nu\sim \nu^{-0.5}$ behaviour.
The complete description of the radio AG requires the inclusion of
two additional effects that, in fair approximations,
introduce no extra parameters:  a ``cumulation factor''
for the electrons that emit the observed radio frequencies
(it takes time for the ISM electrons gathered by the CB to cool to
radio-emitting energies) and an ``illumination and limb-darkening''
factor taking into account that the CBs are viewed relativistically
(an observer would ``see'' almost all of the $4\pi$ surface of a spherical CB).
With these corrections to Eq. (6), the measured broad-band AG of all GRBs
with known z
are well fitted, in spite of the scintillations in the radio.
The overall successful fits [4] involve only one additional
``radio'' parameter, $\nu_a$.
The most complete broad-band data are perhaps those of GRB991208.
Fig. 7b compares its measured radio light-curve at 8.46 GHz
and  the CB model light-curve, obtained from the broad-band fit.
A comparison between its observed and predicted
spectrum between 5 and 10 days after burst is shown in Fig. 7a.

\begin{figure}[htb]
\vspace{-0.1cm}
\centering
\psfig{file=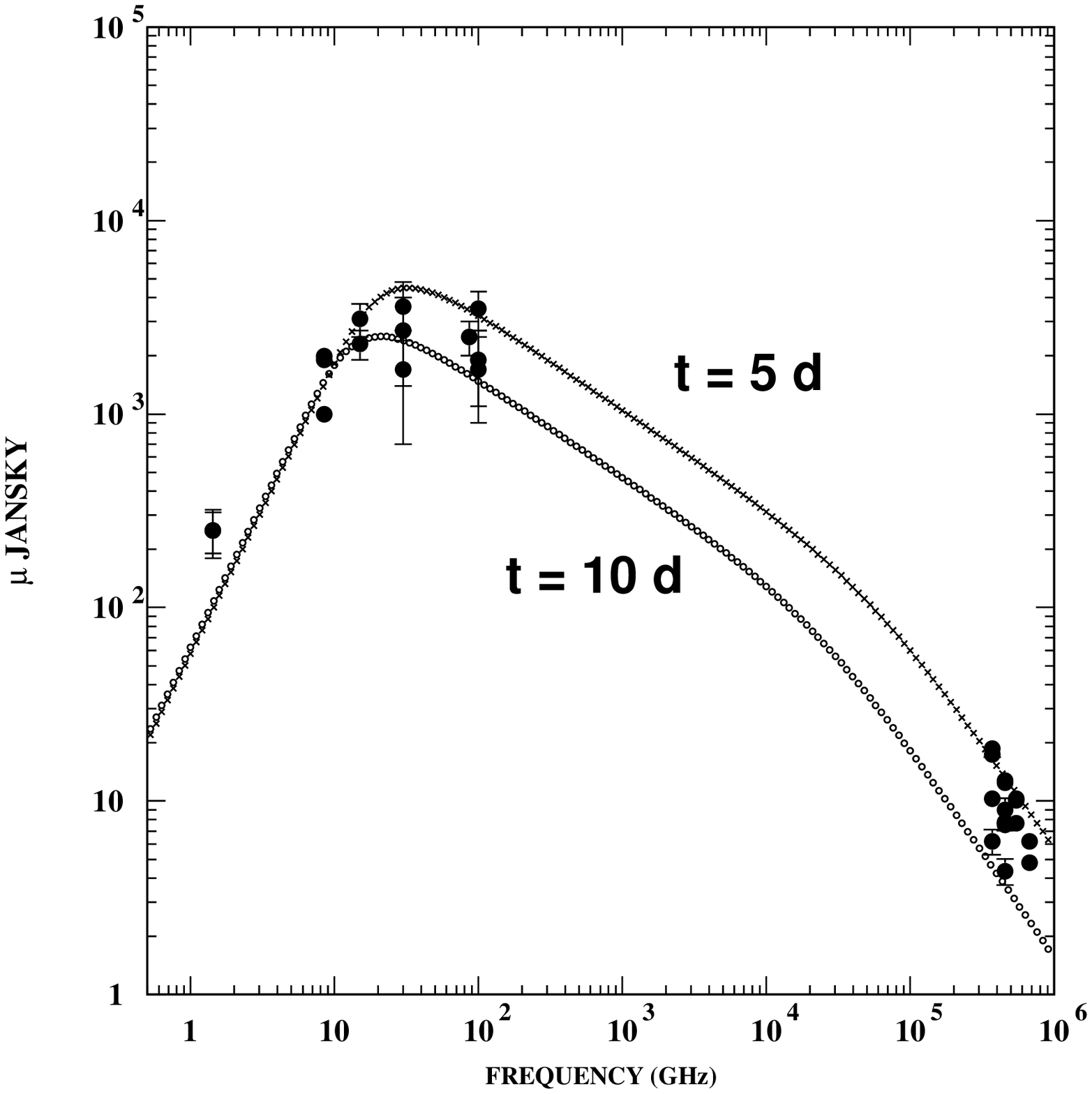,width=6.6cm}\psfig{file=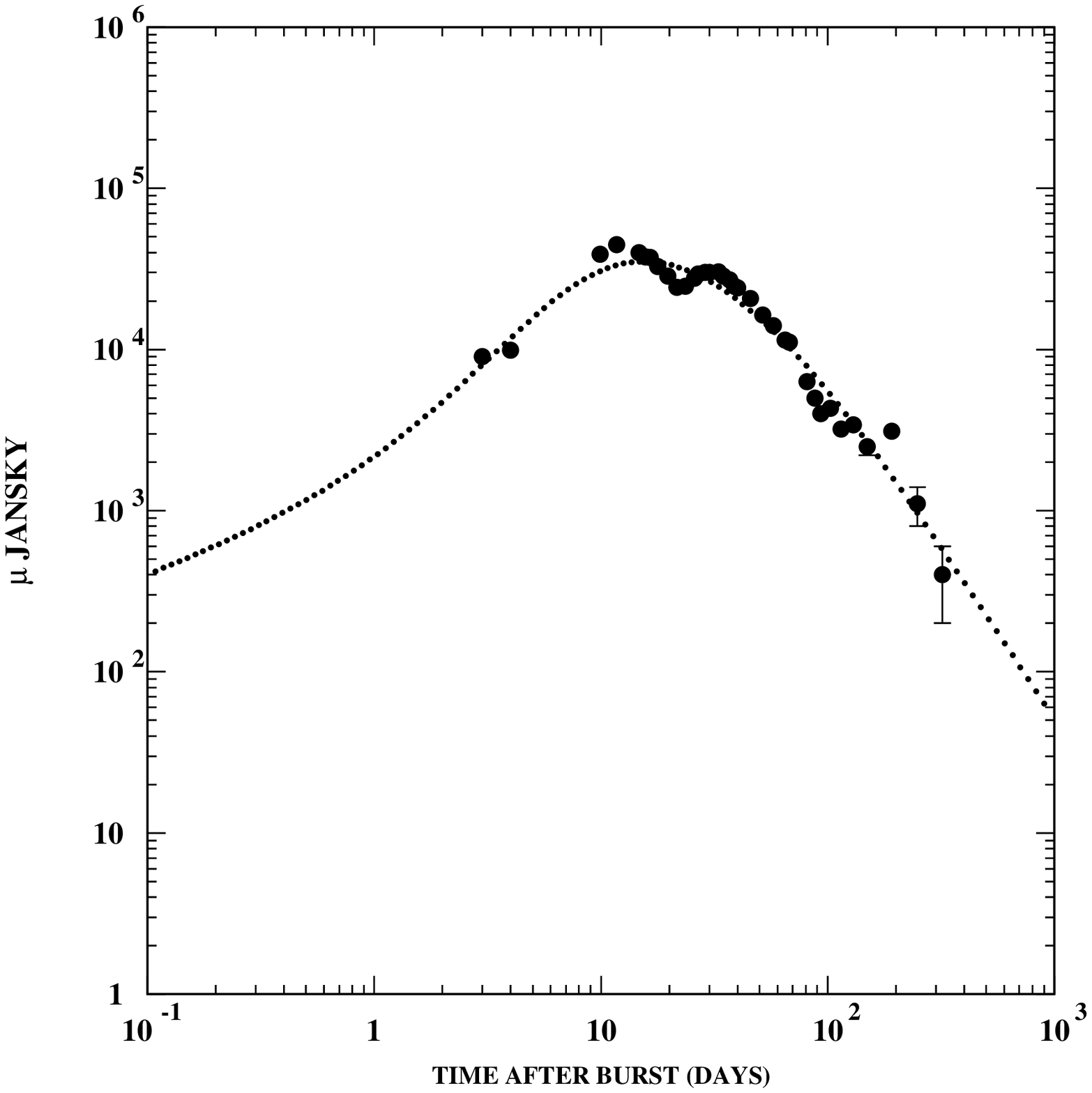,width=6.6cm}
\caption{(a) The CB-model fit to the broad-band spectrum of GRB991208 at
$t=5$ to 10 days. (b) 4.8 GHz light curve of  GRB980425. At other times
and frequencies the fits are equally good.}
\end{figure}

\noindent
{\bf GRB980425/SN1998bw.}
In the CB model, this GRB and its associated
SN1998bw are {\bf not} exceptional. Because it was viewed at
an exceptionally large angle, $\sim 8$ mrad,
its $\gamma$-ray fluence was comparable to that of more distant
GRBs, viewed at  $\theta\sim 1$ mrad [1,3].
That is why its optical AG
was dominated by the SN, except perhaps for the last measured point.
The X-ray AG of its single CB (see Fig. 8a)
is of ``normal'' magnitude, it is {\it not} emitted by the SN.
\begin{figure}[htb]
\vspace{-0.1cm}
\centering
\psfig{file=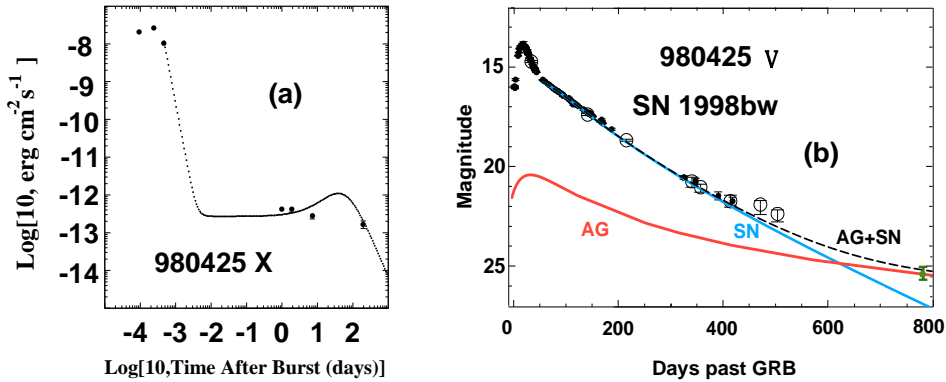,width=12.5cm}
\vspace{-0.3cm}
\caption{CB-model fits to GRB 980425. (a) The X-ray AG. (b) The V-band AG:
the SN contribution, the CB's contribution and the total. All parameters
(but $z$ and $\theta$) are ``normal''. }
\end{figure}
Its predicted late-time behaviour [3]
is consistent with the Chandra and
XMM-Newton measurements reported in this meeting by Kouveliotou and Pian.
The normalization, spectral and temporal behaviour
of the radio AG of this GRB are also ``normal'' and due to
the CB, {\bf not} the SN [3].
The predicted radio light-curve at 4.8 GHz of GRB980425
and the data are shown in Fig.~7b.
SN1998bw, deprived of its ``abnormal''
X-ray and radio emissions (which it did not emit!), loses most of
its ``peculiarity''.

\noindent
{\bf Radio scintillations} of pulsars have been used
to measure their sky-projected velocities, in
agreement with proper-motion measurements.
For cosmological GRBs, the sky angular velocity of their
CBs happens to be comparable to that of the much
slower and closer Galactic pulsars. Perhaps, then,
the analysis of GRB radio oscillations may result in a
 measurement of their apparent
``hyperluminal'' velocities [4].

\noindent
{\bf X-ray lines} observed in the  AG of some GRBs, if real,  may
be Balmer and Lyman lines from hydrogen recombination in the CBs,
Doppler-boosted by their highly relativistic motion and redshifted by the
cosmic
expansion. Then, these lines should be narrow and their observed energy,
$ E(t)=\delta(t)\, E_{line}/(1+z)\, ,$ where
$ E_{line}$ is their energy in the CB rest frame,
should move with time
to lower energy as the CBs decelerate and their Doppler factor
diminishes with time.  Current data are not precise enough to distinguish
between metal lines from photoionized circumburst matter and the CB model
interpretation, but the time-dependence of these lines may be observable
[24].

For lack of time and space I could not discuss the
CB model interpretation of {\it dark bursts, short GRBs and X-ray
Flashes} that  will be published elsewhere.

{\bf Concluding remarks} The CB model is very modest in the adjectives
that refer to GRBs. None of them is exceptional, not even the very
energetic GRB 990123, nor 970508 with its peculiar AG shape, nor the
extraordinarily close-by 980425. They are all associated with asymmetric
supernovae seen from near their axis and visible when they are not too far
or too extinct. The explosions that generate GRBs are neither ``the
biggest
after the Big Bang'' nor ``hypernovae''. The mechanism that begets GRBs is
common: it takes place in quasars as well as  microquasars. The
model
works very well and is very predictive, thus falsifiable.

{\bf acknowledgements}
The author would like to thank the conference organizers Franco
Giovannelli and Giampaolo Mannocchi for the invitation, the theory
division of CERN for its hospitality and the VPR fund for promotion of
research at the Technion for its support.

\end{document}